\documentclass[11pt]{article}
\begin{document}

\title{\bf {Particle creation in an oscillating spherical cavity }}
\author{
M. R. Setare $^1$ \footnote{E-mail: Mreza@physics.sharif.ac.ir}
 and A. A. Saharian  $^{2}$\footnote{E-mail: saharyan@www.physdep.r.am }  \\
 {$^1$ Department of Physics, Sharif University of
Technology, Tehran, Iran}\\ and \\
{$^{2}$}Department of Physics, Yerevan State University, Yerevan,
Armenia }
\date{\small{\today}}
 \maketitle
\begin{abstract}
We study the creation of massless scalar particles from the
quantum vacuum due to the dynamical Casimir effect by spherical
shell with oscillating radius. In the case of a small amplitude of
the oscillation, to solve the infinite set of coupled differential
equations for the instantaneous basis expansion coefficients we
use the method based on the time-dependent perturbation theory of
the quantum mechanics. To the first order of the amplitude we
derive the expressions for the number of the created particles for
both parametric resonance and non-resonance cases.
 \end{abstract}

\newpage

\section{Introduction}
   The Casimir effect is one of the most interesting manifestations
  of nontrivial properties of the vacuum state in a quantum field
  theory (for a review see \cite{mueller,Moste,Milt,Birrell}) and
   can be viewed
  as a polarization of
  vacuum by boundary conditions. A new phenomenon, a
  quantum creation of particles (the dynamical Casimir effect)
  occurs when the geometry of the system varies in time. In two
  dimensional spacetime and for conformally invariant fields the
  problem with dynamical boundaries can be mapped to the
  corresponding static problem and hence allows a complete study
  (see \cite{Moste,Birrell} and references therein). In higher
  dimensions the problem is much more complicated and only
  partial results are available. The vacuum stress induced by uniform
  acceleration of a perfectly reflecting plane is considered in
  \cite{Cand}. The corresponding problem for a sphere
  expanding in the four-dimensional spacetime with constant
  acceleration is investigated by Frolov and Serebriany
  \cite{Frol1,Frol2} in the perfectly reflecting case and by Frolov
  and Singh \cite{Frol3} for semi-transparent boundaries. For more
  general cases of motion the problem of particle and
  energy creation is considered on the base of various
  perturbation methods
\cite{Ford}-\cite{Ander}(in the case of plane boundaries for more
  complete list of references see \cite{dod1}).
   It has been
  shown that a gradual accumulation of small changes in the quantum
  state of the field could result in a significant observable
  effect. A new application of the dynamical Casimir effect has
  recently appeared in connection with the suggestion by Schwinger
  \cite{Schwing} that the photon production associated with changes in
  the quantum electrodynamic vacuum state arising from a
  collapsing dielectric bubble could be relevant for
  sonoluminescence.
  For the further developments and discussions of this quantum-vacuum
  approach see \cite{Eber}-\cite{Liber2} and references
  therein.

In our previous paper \cite{set} we studied the problem of
particle creation from the quantum vacuum by a spherical shell
with time-dependent radius. We have considered examples for which
the mean number of particles can be explicitly evaluated in the
adiabatic approximation when the squeezing effect is dominant. In
the present paper the case is considered when the sphere radius
performs oscillations with a small amplitude and the expressions
are derived for the number of created particles to the first order
of the perturbation theory.

\section{Quantum scalar field inside a sphere \\ with time-dependent radius}

Consider a scalar field $\varphi $ satisfying Dirichlet boundary condition
on the surface of a sphere with time-dependent radius $a=a(t)$:
\begin{equation}
\left( \frac{\partial ^{2}}{\partial t^{2}}-\Delta \right) \varphi
(x,t)=0,\qquad \varphi |_{r=a(t)}=0.  \label{fieldeq}
\end{equation}
Following Ref.\cite{set} we expand the corresponding
eigenfunctions for the interior region in a series with respect to
the instantaneous basis:
\begin{eqnarray}
\varphi _{lmn}(x) &=&\sqrt{\frac{2}{a^{3}(t)}}\sum_{k=1}^{\infty
}q_{nk}^{l}(t)\frac{j_{l}(j_{l,k}r/a(t))}{j_{l}^{\prime }(j_{l,k})}%
Y_{lm}(\theta ,\varphi ),  \label{eigenfunction} \\
l &=&0,1,2,...,\quad -l\leq m\leq l,\quad n=1,2,...,
\end{eqnarray}
where $(r,\theta ,\varphi )$ are standard spherical coordinates,  $j_{l,k}$
is the $n$-th zero for the spherical Bessel function $j_{l}(z)$, $%
j_{l}(j_{l,n})=0$, $Y_{lm}(\theta ,\varphi )$ the spherical harmonic.
These functions automatically satisfy boundary condition (\ref{fieldeq}).
Substituting (\ref{eigenfunction}) into field equation (\ref{fieldeq}) we
arrive at an infinite set of coupled differential equations \cite{set}
\begin{equation}
\ddot{q}_{nk}^{l}+\omega _{lk}^{2}(t)q_{nk}^{l}=2h\sum_{p=1}^{\infty }\dot{q}%
_{np}^{l}a_{pk}^{l}+\dot{h}\sum_{p=1}^{\infty
}q_{np}^{l}a_{pk}^{l}+h^{2}\sum_{p,j=1}^{\infty
}q_{np}^{l}a_{pj}^{l}a_{kj}^{l},  \label{eqcoef}
\end{equation}
where overdot stands for the time derivative,
\begin{equation}
h=\frac{\dot{a}}{a},\quad \omega _{lk}(t)=\frac{j_{l,k}}{a(t)},  \label{hom}
\end{equation}
and the time dependent antisymmetric coefficients $a_{nk}^{l}$ are
determined as
\begin{equation}
a_{nk}^{l}=\left\{
\begin{array}{cc}
0, & \quad k=n \\
2j_{l,n}j_{l,k}/(j_{l,n}^{2}-j_{l,k}^{2}) & \quad k\neq n
\end{array}
\right. .  \label{alnk}
\end{equation}
If the sphere is asymptotically static at past and future then the in- and
out- vacuum states can be defined by using the solutions for coefficients
corresponding to the in- and out-modes $\varphi _{lmn}^{\mathrm{(in)}%
}(t),\,\varphi _{lmn}^{\mathrm{(out)}}(t)$ with asymptotics
\begin{eqnarray}
q_{nk}^{{\mathrm{(in)}}l}(t) &\rightarrow &\frac{e^{-\imath \omega _{ln}^{%
{\mathrm{in}}}t}}{\sqrt{2\omega _{ln}^{{\mathrm{in}}}}}\delta _{nk},\quad
t\rightarrow -\infty   \label{qinout} \\
q_{nk}^{{\mathrm{(out)}}l}(t) &\rightarrow &\frac{e^{-\imath \omega _{ln}^{%
{\mathrm{out}}}t}}{\sqrt{2\omega _{ln}^{{\mathrm{out}}}}}\delta _{nk},\quad
t\rightarrow +\infty ,
\end{eqnarray}
where we use the notations
\begin{equation}
\omega _{ln}^{{\mathrm{in}}}=\frac{j_{l,n}}{a_{-}},\qquad \omega _{ln}^{%
{\mathrm{out}}}=\frac{j_{l,n}}{a_{+}},\qquad a_{\pm }=\lim_{t\rightarrow \pm
\infty }a(t)  \label{omegalf}
\end{equation}
for the corresponding eigenfrequencies. The expansion coefficients for the
in- and out-modes are related by the transformation \cite{set}
\begin{equation}
q_{nk}^{{\mathrm{(in)}}l}=\sum_{j}\left( \alpha _{nj}^{l}
q_{jk}^{{\mathrm{(out)}}%
l}+\beta _{nj}^{l}q_{jk}^{{\mathrm{(out)}}l\ast }\right) ,  \label{qinoutexp}
\end{equation}
where $\alpha _{nj}^{l}$ and $\beta _{nj}^{l}$ are Bogoliubov coefficients
for the Bogoliubov transformation between the in- and out-modes $\varphi
_{lmn}^{{\mathrm{(in)}}}(t)$ and $\,\varphi _{lmn}^{{\mathrm (out)}}(t)$. In
particular, taking into account (\ref{qinout}), in the limit $t\rightarrow
+\infty $ from (\ref{qinoutexp}) we receive
\begin{equation}
q_{nk}^{{\mathrm{(in)}}l}=\frac{1}{\sqrt{2\omega _{lk}^{{\mathrm{out}}}}}
\left(
\alpha _{nk}^{l}e^{-\imath \omega _{lk}^{{\mathrm{out}}}t}+\beta
_{nk}^{l}e^{\imath \omega _{lk}^{{\mathrm{out}}}t}\right) .  \label{qinoutas}
\end{equation}
The mean number of out-particles produced in a given mode with
quantum numbers $l,m,n$ in the
in-vacuum state is determined by the Bogoliubov coefficient $\beta _{nk}^{l}$%
:
\begin{equation}
<{\mathrm{in}}|N_{lmn}|{\mathrm{in}}>=\sum_{k=1}^{\infty }|
\beta _{nk}^{l}|^{2},
\label{Nlmn}
\end{equation}
and the total number of created scalar particles is obtained by
taking the sum over all the modes
\begin{equation}
<{\mathrm{in}}|N|{\mathrm{in}}>=\sum_{l=0}^{\infty }(2l+1)
\sum_{n,k=1}^{\infty
}|\beta _{nk}^{l}|^{2}.  \label{Ntot}
\end{equation}
From the form of equations (\ref{eqcoef}) it follows that there are two
types of effects which lead to the particle creation (see also \cite{Schut}%
). The first one, called squeezing of the vacuum, is due to the
nonstationary eigenfrequencies $\omega _{ln}(t)$ as a result of a dynamical
change of the radius of the sphere and is described by the second term on
the left of (\ref{eqcoef}). The second one, referred as acceleration effect,
is due to the motion of the boundary and comes from the terms on the right
of (\ref{eqcoef}). In \cite{set} we have considered the particle creation
inside a sphere in the adiabatic approximation when the squeezing effect is
dominant. In the next section we will consider another approximation.

\section{Particle creation by an oscillating sphere}

Below we will consider the case when the sphere radius performs small
oscillations at a frequency $\Omega $ during a period $T$:
\begin{equation}
a(t)=\left\{
\begin{array}{cc}
a_{0}[1+\varepsilon \sin (\Omega t)], & \quad 0\leq t\leq T \\
a_{0}, & \quad t<0,\;t>T
\end{array}
\right. ,  \label{radeq}
\end{equation}
where $a_{0}$ is the mean radius, $\varepsilon $ is a small parameter. For
this type of motion $\omega _{ln}^{{\mathrm{in}}}=
\omega _{ln}^{{\mathrm{out}}%
}=\omega _{ln}^{(0)}\equiv j_{l,n}/a_{0}$. We will assume that for $t<0$ the
field is in the in-vacuum state. This means that we need to solve Eq.(%
\ref{eqcoef}) with the initial condition
\begin{equation}
q_{nk}^{l}(t)=\frac{e^{-\imath \omega _{ln}^{(0)}t}}{\sqrt{2\omega
_{ln}^{(0)}}}\delta _{nk},\quad t\leq 0.  \label{initcond}
\end{equation}

To derive the particle number created we will follow the scheme developed in
\cite{Ji} for the case of a perfect plane cavity with vibrating walls.
First of all we note that introducing new functions
\begin{equation}
X_{nk\pm }^{l}(t)=\sqrt{\frac{\omega _{lk}^{(0)}}{2}}\left( q_{nk}^{l}\mp i%
\frac{\dot{q}_{nk}^{l}}{\omega _{lk}^{(0)}}\right) ,  \label{newX}
\end{equation}
to the first order of $\varepsilon $ the set of equations (\ref{eqcoef}) can
be replaced by a coupled first-order differential equations
\begin{eqnarray}
\dot{X}_{nk\pm }^{l} &=&\pm i\omega _{lk}^{(0)}X_{nk\pm }^{l}\mp i\omega
_{lk}^{(0)}\varepsilon \sin (\Omega t)\left( X_{nk+}^{l}+X_{nk-}^{l}\right)
\label{eqcoef4} \\
& & \pm \frac{\varepsilon \Omega }{\sqrt{\omega _{lk}^{(0)}}}\cos
(\Omega t)\sum_{j}a_{jk}^{l}\sqrt{\omega _{lj}^{(0)}}\left(
X_{nj+}^{l}-X_{nj-}^{l}\right) \nonumber \\
& & \pm \frac{i\varepsilon \Omega ^{2}}{\sqrt{%
\omega _{lk}^{(0)}}}\sin (\Omega t)\sum_{j}\frac{a_{jk}^{l}}{\sqrt{\omega
_{lj}^{(0)}}}\left( X_{nj+}^{l}+X_{nj-}^{l}\right) .  \nonumber
\end{eqnarray}
Introducing the column vector
\begin{equation}
{\vec{X}}_{n}^{l}(t)=\left(
\begin{array}{c}
X_{n1-}^{l} \\
X_{n1+}^{l} \\
X_{n2-}^{l} \\
\vdots
\end{array}
\right)   \label{Xvec}
\end{equation}
this system can be written in the matrix form
\begin{equation}
\frac{d}{dt}{\vec{X}}_{n}^{l}(t)=V^{(0)}{\vec{X}}_{n}^{l}(t)+\varepsilon
V^{(1)}{\vec{X}}_{n}^{l}(t).  \label{matrixeq}
\end{equation}
Here the components for the matrices $V^{(0)}$ and $V^{(1)}$ are
\begin{equation}
V_{k\sigma ,j\sigma ^{\prime }}^{(0)}=i\omega _{lk}^{(0)}\sigma \delta
_{kj}\delta _{\sigma \sigma ^{\prime }},\quad V_{k\sigma ,j\sigma ^{\prime
}}^{(1)}=\sum_{s=\pm }v_{k\sigma ,j\sigma ^{\prime }}^{s}e^{is\Omega t},
\label{V0V1}
\end{equation}
where
\begin{equation}
v_{k\sigma ,j\sigma ^{\prime }}^{s}=-\frac{1}{2}s\sigma \omega
_{lk}^{(0)}\delta _{kj}+\sigma \Omega a_{jk}^{l}\sqrt{\frac{\omega
_{lj}^{(0)}}{\omega _{lk}^{(0)}}}\left( \frac{\sigma ^{\prime }}{2}+\frac{%
s\Omega }{4\omega _{lj}^{(0)}}\right) ,  \label{vs}
\end{equation}
and $s,\sigma ,\sigma ^{\prime }=+,-$. To solve Eq. (\ref{matrixeq}) we use
a perturbation expansion
\begin{equation}
{\vec{X}}_{n}^{l}={\vec{X}}_{n}^{(0)l}+\varepsilon {\vec{X}}%
_{n}^{(1)l}+\cdots ,  \label{pertexp}
\end{equation}
where at the zeroth order, by taking into account initial condition (\ref
{initcond}), one has
\begin{equation}
X_{nk\sigma }^{(0)l}=\delta _{nk}\delta _{\sigma -}e^{-i\omega _{lk}^{(0)}t}.
\label{Xl0}
\end{equation}
The first order term can be easily found on the base of this expression:
\begin{equation}
X_{nk\sigma }^{(1)l}=e^{i\sigma \omega _{lk}^{(0)}t}\int_{0}^{t}dt^{\prime
}e^{-i\sigma \omega _{lk}^{(0)}t^{\prime }}\sum_{j,\sigma ^{\prime
}}V_{k\sigma ,j\sigma ^{\prime }}^{(1)}X_{nk\sigma ^{\prime }}^{(0)l}.
\label{Xl1}
\end{equation}
Using (\ref{Xl0}) and (\ref{V0V1}) one finds
\begin{equation}
X_{nk\pm }^{(1)l}(t) =\mp v_{k-,n-}^{-}E_{\pm \omega _{lk}^{(0)}+\Omega
+\omega _{ln}^{(0)}}^{\mp \omega _{lk}^{(0)}}\mp v_{k-,n-}^{+}E_{\pm \omega
_{lk}^{(0)}-\Omega +\omega _{ln}^{(0)}}^{\mp \omega _{lk}^{(0)}}
\label{Xlpm}
\end{equation}
where we use the notation
\begin{equation}
E_{m}^{\omega _{lk}^{(0)}}(t)=\left\{
\begin{array}{cc}
te^{-i\omega _{lk}^{(0)}t}, & \quad \mathrm{for}\quad m=0 \\
(i/m)\left[ e^{-i(m+\omega _{lk}^{(0)})t}-e^{-i\omega _{lk}^{(0)}t}\right] ,
& \quad \mathrm{for}\quad m\neq 0
\end{array}
\right. .  \label{Em}
\end{equation}
For the instantaneous basis expansion coefficients this yields
\begin{eqnarray}
q_{nk}^{l}(t) &=&\frac{e^{-i\omega _{lk}^{(0)}t}}{\sqrt{2\omega _{lk}^{(0)}}}%
\delta _{nk}+\frac{\varepsilon }{\sqrt{2\omega _{lk}^{(0)}}}\left\{
v_{k-,n-}^{-}E_{-\omega _{lk}^{(0)}+\Omega +\omega _{\ln }^{(0)}}^{\omega
_{lk}^{(0)}}+v_{k-,n-}^{+}E_{-\omega _{lk}^{(0)}-\Omega +\omega
_{ln}^{(0)}}^{\omega _{lk}^{(0)}}\right.   \label{qlnk1ord} \\
&-&\left. v_{k-,n-}^{-}E_{\omega _{lk}^{(0)}+\Omega +\omega
_{ln}^{(0)}}^{-\omega _{lk}^{(0)}}-v_{k-,n-}^{+}E_{\omega _{lk}^{(0)}-\Omega
+\omega _{ln}^{(0)}}^{-\omega _{lk}^{(0)}}\right\} +\mathcal{O}(\varepsilon
^{2}).  \nonumber
\end{eqnarray}
This expression includes terms proportional to $t$ which are due to the
parametric resonance. In the situation $\omega _{lk}^{(0)}t\gg 1$ the
resonance terms are dominant and solution (\ref{qlnk1ord}) becomes
\begin{eqnarray}
q_{nk}^{l}(t) &\approx &\frac{e^{-i\omega _{lk}^{(0)}t}}{\sqrt{2\omega
_{lk}^{(0)}}}\delta _{nk}+\frac{\varepsilon t}{\sqrt{2\omega _{lk}^{(0)}}}%
\left[ -v_{k-,n-}^{+}e^{i\omega _{lk}^{(0)}t}\delta _{\omega
_{lk}^{(0)},\Omega -\omega _{ln}^{(0)}}\right.   \label{qlnkpar} \\
&+&\left. v_{k-,n-}^{-}e^{-i\omega _{lk}^{(0)}t}\delta _{\omega
_{lk}^{(0)},\Omega +\omega _{ln}^{(0)}}+v_{k-,n-}^{+}e^{-i\omega
_{lk}^{(0)}t}\delta _{\omega _{ln}^{(0)},\Omega +\omega _{lk}^{(0)}}\right] .
\nonumber
\end{eqnarray}
After the time interval $T$ the sphere is static with radius $a_{0}$ and the
function $q_{nk}^{l}(t)$ has the form (\ref{qinoutas}). From the continuity
condition between the functions (\ref{qinoutas}) and (\ref{qlnkpar}) at $t=T$
one has
\begin{eqnarray}
\alpha _{nk}^{l} &=&\delta _{nk}+\varepsilon T\left[ v_{k-,n-}^{-}\delta
_{\omega _{lk}^{(0)},\Omega +\omega _{ln}^{(0)}}+v_{k-,n-}^{+}\delta
_{\omega _{ln}^{(0)},\Omega +\omega _{lk}^{(0)}}\right]   \label{Bogalf} \\
\beta _{nk}^{l} &=&-\varepsilon Tv_{k-,n-}^{+}\delta _{\omega
_{lk}^{(0)},\Omega -\omega _{ln}^{(0)}}.  \label{Bogbet}
\end{eqnarray}
By taking into account expressions (\ref{vs}) and (\ref{alnk}) one
has
\begin{equation}
\beta _{nk}^{l}=-\frac{1}{2}\varepsilon T\sqrt{\omega _{ln}^{(0)}\omega
_{lk}^{(0)}}\delta _{\omega _{lk}^{(0)},\Omega -\omega _{ln}^{(0)}}.
\label{betparres}
\end{equation}
Now from Eq.(\ref{Nlmn}) for the number of out-particles with the energy $%
\omega _{ln}^{(0)}$ and quantum numbers $l,m$ created by the parametric
resonance in the in-vacuum state we receive
\begin{equation}
<{\mathrm{in}}|N_{lmn}|{\mathrm{in}}>=\frac{1}{4}(\varepsilon
T)^{2}\sum_{k=1}^{\infty }\omega _{ln}^{(0)}\omega _{lk}^{(0)}\delta
_{\omega _{lk}^{(0)},\Omega -\omega _{ln}^{(0)}}.  \label{Nlmn2}
\end{equation}
As we see the necessary condition for the parametric resonance is
\begin{equation}
\Omega =\frac{1}{a_{0}}\left( j_{l,q}+j_{l,p}\right)   \label{parrescond}
\end{equation}
for some values $q$ and $p$.

Now we turn to calculating the Bogoliubov coefficient $\beta
^{l}_{nk}$ in the non-resonance case. By using the expressions
(\ref{Em})
from the continuity condition between the functions (\ref{qinoutas}) and (%
\ref{qlnkpar}) and their derivatives at $t=T$ after some algebra we get
\begin{equation}
\beta _{nk}^{l} =-2\varepsilon \exp \left[ -iT\left( \omega
_{ln}^{(0)}+\omega _{lk}^{(0)}\right) /2\right] \sum_{\sigma =+,-}
v_{k-,n-}^{\sigma }e^{i\sigma \Omega T/2}\frac{\sin \left[ T\left( \omega
_{ln}^{(0)}+\omega _{lk}^{(0)}-\sigma \Omega \right) /2\right] }{\omega
_{ln}^{(0)}+\omega _{lk}^{(0)}-\sigma \Omega }   \label{betnonres}
\end{equation}
For the number of quanta we derive from here
\begin{eqnarray}
<{\mathrm{in}}|N_{lmn}|{\mathrm{in}}>&=&4\varepsilon ^{2}\sum_{k=1}^{\infty
}\left\{ \sum_{\sigma =+,-}\left( v_{k-,n-}^{\sigma }\right) ^{2}
\frac{\sin ^{2}\left[ T\left(
\omega _{ln}^{(0)}+\omega _{lk}^{(0)}-\sigma \Omega \right) /2\right] }{\left(
\omega _{ln}^{(0)}+\omega _{lk}^{(0)}-\sigma \Omega \right) ^{2}}\right.
\label{Nlmnnonres} \\
&+&\left. \frac{v_{k-,n-}^{-}v_{k-,n-}^{+}\cos (\Omega T)}{\left( \omega
_{ln}^{(0)}+\omega _{lk}^{(0)}\right) ^{2}-\Omega ^{2}}\left[ \cos (\Omega
T)-\cos \left[ T\left( \omega _{ln}^{(0)}+\omega _{lk}^{(0)}\right) \right] %
\right] \right\} .  \nonumber
\end{eqnarray}
Note that in the resonance case from the term with $\sigma =+$
of Eq.(\ref{betnonres}) we
recover the result (\ref{Bogbet}).

Up to now we have considered the scalar field with the Dirichlet
boundary condition. In a similar way it is easy to generalize the
corresponding results for other type of boundary conditions. For
example, in the case of the Neumann scalar the number of particles
created inside an spherical shell is given by expressions
(\ref{Nlmn2}) and (\ref{Nlmnnonres}) where now $a\omega
^{(0)}_{ln}$, $n=1,2,...$ are zeros for the function $j_{l}'(z)$,
$j_{l}'(a\omega ^{(0)}_{ln})=0$.

\section{Concluding remarks}

Our interest in this paper has been in the quantum radiation by an
spherical mirror with time-dependent radius given by (\ref{radeq})
and under assumption that the amplitude of oscillation is small.
To define and interpret asymptotic spaces of physical states
uniquely we assumed that the oscillation of the sphere lasts a
finite period of time. We consider a massless scalar field obeying
the Dirichlet boundary condition on the sphere surface. To solve
the corresponding equations for the spontaneous basis expansion
coefficients in the short-time limit, $\varepsilon \Omega T\ll 1$,
we use the method developed in \cite{Ji} for the case of plane
boundaries. To the first order in perturbation theory we derive
the expression for the number of particles created inside the
sphere in both resonance and non-resonance cases. In the resonance
case the oscillation frequency is given by expression
(\ref{parrescond}). The corresponding solution includes resonance
terms proportional to $t$ and oscillating parts. In the situation,
when $\Omega T\gg 1$ the resonance terms are dominant, and the
total number of particles created in the mode with quantum numbers
$lmn$ and with frequency $\omega _{ln}^{(0)}$ is given by
(\ref{Nlmn2}). In this case the number of particles grows
quadratically in time. In the non-resonance case the corresponding
formula for the number of particles has the form
(\ref{Nlmnnonres}). The generalization of these formulas for the
other boundary conditions is straightforward. Another possible
generalization is to perform the higher order calculations with
respect to $\varepsilon $. We hope that the results obtained in
present paper may have an application in studies of quantum
radiation created by bubble formation during first order phase
transitions in the Early Universe.

\end{document}